\documentclass[aps,pra,twocolumn,superscriptaddress]{revtex4}

\usepackage{amsmath}
\usepackage{amsfonts}
\usepackage{amssymb}
\usepackage{epsfig}
\usepackage{graphicx}
\usepackage{wrapfig}
\usepackage{bm}
\usepackage{upgreek} 
\usepackage{cancel} 
\usepackage[caption=false]{subfig}
\usepackage{float}
\usepackage[utf8x]{inputenc}
\usepackage[usenames]{color}
\usepackage{multirow}
\usepackage{graphicx}
\usepackage{hyperref}
\usepackage{tikz}
\usepackage{3dplot}

\graphicspath{{figures/}} 

\renewcommand{\vr}{\mathbf{r}}

\begin{document}

\preprint{draft version}

\title{Quantum Entanglement in (d-1)-Spherium}

\author{I. V. Toranzo$^{1}$, A. R. Plastino$^{1,2}$, P. S\'anchez-Moreno$^{1,3}$
and J. S. Dehesa$^{}$}

\address{
  $^{1}$Instituto Carlos I de F\'{\i}sica Te\'orica y
Computacional and Departamento de F\'{\i}sica At\'omica Molecular
y Nuclear, Universidad de Granada, Granada 18071, Spain.\\
  $^{2}$CeBio y Secretar\'{\i}a de Investigaci\'on, Universidad Nacional
  Noroeste-Buenos Aires, UNNOBA-Conicet, Roque Saenz Pe\~na 456, Junin, Argentina.\\
$^{3}$Departamento de Matemática Aplicada, Universidad de Granada, Granada 18071, Spain.}

\date{\today}

\begin{abstract}
There are very few systems of interacting particles (with continuous variables)
for which the entanglement of the concomitant eigenfunctions can be computed in
an exact, analytical way. Here we present analytical calculations of the amount of
entanglement exhibited by $s$-states of {\it spherium}. This is a system of two
particles (electrons) interacting via a Coulomb potential and confined to a
$(d-1)$-sphere (that is, to the surface of a $d$-dimensional ball). We investigate
the dependence of entanglement on the radius $R$ of the system, on the spatial
dimensionality $d$, and on energy. We find that entanglement increases monotonically with $R$,
decreases with $d$, and also tends to increase with the energy of the eigenstates.
These trends are discussed and compared with those observed in other two-electron
atomic-like models where entanglement has been investigated.

\vskip
0.5cm

\noindent
Pacs: 03.65.-w, 03.67.-a, 03.67.Mn

\end{abstract}

\maketitle

\section{Introduction}
It has been recently shown by Loos and Gill \cite{LG09,LG10}
that ``spherium", a system consisting of two electrons trapped on the
surface of a sphere and interacting through a Coulomb potential,
belongs to the family of quasi-exactly solvable quantum mechanical models.
These are models whose Schr\"odinger eigenvalue equation admits
an explicit analytical solution for a finite portion of the energy
spectrum. This kind of models are of considerable interest both for
illuminating the properties of more complex or realistic systems and for testing
and developing approximate treatments, such as those related to density functional
theory. Indeed, spherium has found interesting applications in the study of
correlated quantum systems (see \cite{LG09} and references therein).
Spherium is related to another widely studied semi-solvable two-body model,
the Hooke atom, which consists of a pair of electrons repelling Coulombically
and confined by a harmonic external potential (this system has direct
experimental relevance as a model of a two-electron quantum dot
with parabolic confinement). Here we are going to consider a
$(d-1)$-dimensional version of spherium, where the two electrons
are trapped on a $(d-1)$-sphere of radius $R$. By a $(d-1)$-sphere we mean
the surface of a $d$-dimensional ball.\\

Exactly solvable and semi-solvable systems provide a valuable arena for
the exploration of the entanglement properties of quantum systems
of interacting particles. In particular, they provide useful insights for illuminating
the entanglement-related features of natural and artificial atomic systems.
Unfortunately, there are few such systems where entanglement measures can be
evaluated analytically. In point of fact, to the best of our knowledge,
the only system of two interacting particles with continuous variables
where entanglement has been calculated in an exact analytical way is the
Moshinsky model \cite{YPD10,BMPSD2012}. Even for the Hooke atom, entanglement calculations are based
upon the numerical evaluation of rather complex multi-dimensional integrals \cite{MPDK10}.\\

In the present contribution we show that spherium is a highly exceptional model,
where the amount of entanglement exhibited by some of its eigenstates
can be determined in an exact and fully analytical way. As far as we know,
spherium is the only two-body system with Coulomb interaction where this
goal has been achieved.\\

Entanglement is nowadays regarded as one of the most fundamental phenomena
in Quantum Physics \cite{BZ06,AFOV08,TMB10}.
Entangled states of multipartite quantum systems are endowed with non-classical
correlations that give rise to a variegated family of physical
phenomena of both fundamental and technological significance. Quantum
entanglement can be viewed in two complementary ways.
On the one hand, entanglement constitutes a valuable resource. The controlled
manipulation of entangled states is central to several quantum
information technologies. On the other hand, entanglement can be regarded as
a fundamental ingredient for the physical characterization of natural quantum
systems such as, for instance, atoms and molecules.  These two points
of view are closely related to each other, although the latter is somehow
less developed than the former. Concerning the second of the
approaches mentioned above, several researchers have investigated in recent years the phenomenon
of entanglement in two-electron atomic models and related systems
\cite{YPD10,BMPSD2012,MPDK10,TMB10,AM04,OS07,DKYPE2012,OS08,CSD08,PN09,HF11,KO10,K11,KH2012,KO13,KO14,GN05,MPD2012,NSPC12,
SB12,LH13,SFM13,LLH13,RS14,RS2015,LH14}.
Most works dealing with entanglement in two-electron systems have
been restricted to the associated ground state wavefunctions.
However, the entanglement properties of excited states of
two-electron atomic models have also been investigated \cite{YPD10,MPDK10}.
The most detailed results concerning the entanglement of excited states have
been obtained from analytical investigations of exactly soluble models, in particular
the Moshinsky one \cite{YPD10}.\\

 The main entanglement-related features exhibited by
 these models share some common trends. First, one observes that entanglement increases
 with the strength of the interaction between the particles.
 Alternatively, for a constant interaction strength,
 entanglement decreases with the strength of the confining potential (this behaviour
 has also been verified in numerical studies of entanglement in Helium-like atoms
 with increasing nuclear charge).
 These effects are clearly two sides of the same coin, and can usually
 be described jointly in terms of the dependence of entanglement
 on an appropriate dimensionless parameter corresponding to the
 relative strengths of the interaction and the confining potentials.
 In the case of atomic-like models with an external harmonic confining potential,
 such as the Moshinsky and the Hooke ones, it is also observed that
 entanglement tends to increase with energy. This last property hold for the majority of states.
 However there are a few entanglement ``level-crossings" where a state has more entanglement than
 another state of higher energy \cite{YPD10}. Aside from these rare exceptions, the
 general monotonically increasing behaviour of entanglement with energy has
 been observed in harmonically confined models endowed with different types
 of particle interaction (i.e., harmonic interaction in the Moshinsky system,
 Coulomb interaction in the Hooke atom; and a $r^{-2}$-interaction potential
 in the Crandall model). Another trend exhibited by two-electron models with harmonic confinement (which
 also holds for different interaction laws between the constituent particles) is that
 the amount of entanglement associated with excited states does not always
 vanish in the limit of a vanishing interaction \cite{MPD2012}. \\

The goal of the present paper is to calculate analytically the amount of entanglement of the ground
state of $(d-1)$-spherium.
The paper is organized as follows. In Section II, we briefly review the concept of entanglement
in systems consisting of identical fermions. In Section III, we show the technical details of the calculations performed in this work. In Section IV, we describe our main results. Finally, some conclusions are drawn in Section V.\\

\section{Entanglement in systems of identical fermions}

There is a natural and physically meaningful measure of entanglement
for pure states of systems consisting of two identical fermions. It is
based on the Schmidt decomposition for fermions \cite{NV07,PMD09}, which reads

\begin{equation}
\label{eq:e1}
|\Psi\rangle = \sum_{i} \sqrt{\frac{\lambda_{i}}{2}}\left(|2 i\rangle |2 i +1\rangle - |2 i +1\rangle |2 i\rangle   \right),
\end{equation}

\noindent
 where  $\{|i\rangle , i = 0, 1,\ldots\}$ is an appropriate orthonormal basis
 of the single-particle Hilbert space, and $0\leq \lambda_{i} \leq 1$ with $\sum_{i} \lambda_{i} =1$.
 The entanglement of the pure state $|\Psi\rangle$ can then be expressed
 in terms of the above fermionic Schmidt coefficients, as

\begin{equation}
\label{eq:e2}
\xi[|\Psi\rangle] = 1 - \sum_{i} \lambda_{i}^{2} = 1 - 2\, \text{Tr} (\rho_{1}^{2}),
\end{equation}

\noindent
where $\rho_{1}= Tr_{2} |\Psi\rangle \langle \Psi|$ is the  single-particle reduced
density matrix obtained from the global, two-particle density
matrix $\rho = |\Psi\rangle \langle \Psi|$. The Schmidt coefficients
$\lambda_i$ are the eigenvalues (each one two-fold degenerate) of $\rho_1$.
The entanglement measure (\ref{eq:e2}) is (up to appropriate additive and
multiplicative constants) basically given by the linear entropy
$S_L(\rho_1) = 1- Tr (\rho_1^2)$  of the single-particle density matrix
$\rho_1$. Alternatively, one could consider an entanglement measure based
upon the von Neumann entropy of the density matrix $\rho_1$,
given by $S_{\rm vN}(\rho_1) = -Tr \rho_1 \ln \rho_1$. This last measure
is extremely difficult to evaluate analytically for systems with continuous variables.
Even in the case of the Moshinsky model, which is the atomic model where entanglement
has been studied more systematically \cite{YPD10,BMPSD2012}, the entanglement measure based on the von Neumann
entropy has been determined in an exact analytical way only for the ground state \cite{AM04,PN09}.
It is highly unlikely that in systems with Coulomb interactions the entanglement
measure based on the von Neumannn entropy can be calculated analytically.
In these cases the (exact) analytical approach seems basically intractable.
Entanglement measures based on the linear entropy have many computational
advantages, both from the analytical and the numerical points of view.
In particular, and in contrast with measures based on the von Neumann entropy,
measures based on the linear entropy can be evaluated directly from $\rho_1$,
without the need of first determining $\rho_1$'s eigenvalues. They constitute
a practical tool for assessing the amount of entanglement
that has been applied to the study of a variety
of systems (see \cite{YPD10,BMPSD2012,MPDK10,NV07} and references therein). \\

An important property of the entanglement measure (\ref{eq:e2}) is that correlations
between the two particles that are solely due to the antisymmetry of the
fermionic state do not contribute to the state's entanglement. In fact, the
amount of entanglement exhibited by a two-fermion state is given,
basically, by the quantum correlations that the state has beyond the
 minimum correlations required by the antisymmetric constraint on
 the fermionic wavefunction
 \cite{ESBL02,GM04,GMW02,NV07,OSTS08,BPCP08,ZP10,PMD09}.
 Consequently, the entanglement of a pure state of two identical fermions
 that can be written as a single Slater determinant is zero. \\

 We apply now the above measure to a pure state of a two-electron system.
 In order to analyze the entanglement of the eigenstates of spherium we
 have to consider states described by wavefunctions of the form,

\begin{equation}
\label{eq:e3}
\psi(\vr_{1},\vr_{2})\chi(\sigma_{1},\sigma_{2}),
\end{equation}

\noindent
with the total wavefunction  factorized as the product of a coordinate
wavefunction $\psi(\vr_{1},\vr_{2})$ and a spin wavefunction
$\chi(\sigma_{1},\sigma_{2})$. Here $\vr_{1}$ and $\vr_{2}$ stand
for the vector positions of the two electrons. The density matrix
corresponding to a wavefunction of the form (\ref{eq:e3}) is given by

\begin{equation}
\label{eq:e4a}
\rho = \rho^{(coord.)} \otimes \rho^{(spin)}
\end{equation}

\noindent
where the matrix elements of $\rho^{(coord.)}$ are

\begin{equation}
\label{eq:e4}
\langle \vr_{1}' , \vr_{2}' | \rho^{(coord.)} |  \vr_{1} , \vr_{2} \rangle = \psi(\vr_{1}' , \vr_{2}') \psi^{*}(\vr_{1} , \vr_{2}).
\end{equation}

\noindent
Even if we are going to investigate the entanglement features only of pure states
of spherium, it is conceptually convenient to consider the corresponding
density matrix (proyector) (\ref{eq:e4a}) in order to obtain from it
the single-particle reduced density matrix, in terms of
which the entanglement measure to be used can be clearly formulated.
For a state with a wavefunction of the
form (\ref{eq:e3}) (and a density matrix of the form (\ref{eq:e4a}))
the entanglement measure (\ref{eq:e2}) reads,

\begin{eqnarray}
\label{eq:e45}
\xi[|\Psi\rangle] &=& 1- 2 \text{Tr} \left[\rho_1^2 \right] \nonumber  \\
&=& 1 - 2 \, \text{Tr} \left[\left( \rho_{1}^{(coord.)}  \right)^{2}  \right] \text{Tr} \left[\left(\rho_{1}^{(spin)}   \right)^{2} \right],
\end{eqnarray}

\noindent
where $\rho_1 = \rho_{1}^{(coord.)} \otimes \rho_{1}^{(spin)}$ is the single-particle
reduced density matrix, and $\rho_{1}^{(coord.)}$ and $\rho_{1}^{(spin)}$ are,
respectively, the marginal density matrices obtained after tracing the matrices $\rho^{(coord.)}$ and $\rho^{(spin)}$ over the degrees of freedom of one of the two particles. It is clear that the entanglement between the two electrons described by (\ref{eq:e3}) involves both the translational and the spin degrees of freedom of electrons.\\

To calculate the entanglement measure (\ref{eq:e45}), it is necessary to consider separately the cases of a spin wavefunction corresponding to parallel spins or antiparallel spins. When spins are parallel (that is, when the coordinate wavefunction is antisymmetric and the spin wavefunction is either $\chi_{++}$ or $\chi_{--}$), one has $Tr[(\rho^{(spin)})^{2}] = 1$, and the entanglement
 measure (\ref{eq:e45}) of a two-electron state of the form (\ref{eq:e3}) is

\begin{equation}
\label{eq:e5}
\xi[|\Psi\rangle] =  1 - 2 \int |\langle \vr_{1}' |\rho_{1}^{(coord.)} | \vr_{1} \rangle |^{2} \, d\vr_{1}' d\vr_{1}
\end{equation}

\noindent
On the other hand, when the spins are anti-parallel (when the coordinate wavefunction is symmetric and the spin wavefunction is $\frac{1}{\sqrt{2}}(\chi_{+-} - \chi_{-+})$, or alternatively, when the coordinate wavefunction is antisymmetric and the spin wavefunction is $\frac{1}{\sqrt{2}}(\chi_{+-} + \chi_{-+})$), one has $Tr [(\rho^{(spin)})^{2}]=\frac{1}{2}$, and the entanglement is

\begin{equation}
\label{eq:e6}
\xi[|\Psi\rangle] =  1 -  \int |\langle \vr_{1}' |\rho_{1}^{(coord.)} | \vr_{1} \rangle |^{2} \, d\vr_{1}' d\vr_{1}
\end{equation}

\noindent
In equations (\ref{eq:e5}) and (\ref{eq:e6}) we have

\begin{equation}
\label{eq:e7}
\langle \vr_{1}' |\rho_{1}^{(coord.)} |\vr_{1} \rangle = \int \psi(\vr_{1}', \vr_{2}) \psi^{*}(\vr_{1}, \vr_{2})\, d\vr_{2}
\end{equation}
for the matrix elements of the \textit{coordinate} marginal matrix density.\\

In the above discussion we have considered two-electron states that are
separable with respect to the spin and spatial degrees of freedom.
Moreover, among these states we only considered states where the spin parts
of the wavefunction correspond to the standard singlet and triplet states.
When studying two-electron systems with a Hamiltonian not depending on
spin, and energy levels with no degeneracy arising from the spatial
part of the Hamiltonian, it is standard and natural to focus on
eigenstates of the above described forms. However, even in these
cases the spin-independence of the Hamiltonian leads to degeneracy of
the energy spectra, and to the existence of eigenstates with the spin
part of the wavefunction different from the ones just mentioned.
For instance, one can have as spin wavefunction a linear combination
of the triplet states. The corresponding (global) eigenstate would have
an amount of entanglement different from the ones given by equations
(\ref{eq:e5}-\ref{eq:e6}). But the difference would be due solely to the spin part,
and would not correspond to any specific feature of the particular
two-electron system under consideration. If the Hamiltonian includes
spin-orbit interaction terms, coupling the spin and the spatial degrees of freedom,
the situation becomes much more complex. The eigenstates would not, in general, have
the spin and the spatial parts disentangled. In such cases both types of degrees
 of freedom need to be considered jointly in order to evaluate the entanglement
 between the two electrons constituting the system. These situations are outside
 the scope of the present work. The spherium Hamiltonian does not depend on spin,
 and we shall consider only $s$-states, where the spatial part of the wavefunction
 is symmetric, and the spin part is given by the singlet state.

\section{$(d-1)$-spherium: description}

As already mentioned, spherium consists of two identical
particles (``electrons") interacting via a Coulomb potential
and confined to the surface of a $(d-1)$-sphere of radius $R$. The corresponding Hamiltonian,
expressed in atomic units, reads,

\begin{equation}
\label{hamilto}
H = -\frac{\nabla^2_1}{2} -\frac{\nabla^2_2}{2} + \frac{1}{r_{12}},
\end{equation}

\noindent
where $r_{12} = |\vr_{1} - \vr_{2}|$ is the interelectronic
distance (a brief review of some basic aspects of the spherium Hamiltonian
is given in Appendix A).
$^{1}S$ states ($s$-states) have a wavefunction $\Psi(r_{12})$ that depends
only on the inter-electronic distance. The corresponding Schr\"odinger
equation can be cast in the form,

\begin{equation}
\label{schroe}
\left[\frac{u^2}{4R^2} -1\right] \frac{d^2\Psi}{du^2} +
\left[\frac{u(2d-3)}{4R^2} - \frac{d}{u}  \right] \frac{d\Psi}{du}
+\frac{\Psi}{u} = E \Psi,
\end{equation}

\noindent
where $u = r_{12}$. As was recently proved by Loos and Gill in
\cite{LG09}, equation  (\ref{schroe}) admits closed analytical solutions for
particular, discrete values of the radius $R=R_{n,m}$. These
exact eigenfunctions of the spherium system have a polynomial form,

\begin{equation}
\label{eq:singwav}
\Psi_{n,m}(r_{12}) = \sum_{k=0}^{n} s_{k,m} r_{12}^{k},
\end{equation}

\noindent
where the coefficients $s_{k,m} \equiv s_{k,m}(d)$ are determined by the recurrence relation

\begin{equation}
\label{eq:recuterms}
s_{k+2,m} = \frac{s_{k+1,m}+\left[k(k +2 (d-1)-2)\frac{1}{4 R_{n,m}^{2}} - E_{n,m} \right]s_{k,m}}{(k+2)(k+(d-1))},
\end{equation}

\noindent
with the starting values $s_{0,m} =1$ and $s_{1,m}=\frac{1}{(d-1)-1}\equiv \gamma$.
The integer parameter $n$ has values $n=1,2, \ldots$ and $m$ is the number of
roots that the polynomial (\ref{eq:singwav}) has in the range $[0,2R]$. That is,
the wavefunction (\ref{eq:singwav}) corresponds to the $m$-th excited $s$-state.
\\

For a given $n$, the energies are obtained by finding the roots of the equation $s_{n+1,m}=0$,
which is a polynomial in $E$, of degree $(n+1)/2$.
The corresponding radius $R_{n,m}$ is found through the relation

\begin{equation}
\label{eq:radsing}
R_{n,m}^{2}E_{n,m} = \frac{n}{2}\left( \frac{n}{2}+(d-1) -1  \right).
\end{equation}

\noindent
We see that the special $R$-values for which the $s$-states of spherium
can be obtained in a closely analytical way arise from an expansion of
the wavefunction in powers of $r_{12}$ that, for the mentioned $R$-values,
becomes a finite polynomial (for a full discussion see \cite{LG09,LG10}
and references therein). Of course, the spherium system is well defined for any value of $R$, and the corresponding Schr\"odinger equation can be solved numerically, leading to results that
interpolate between those corresponding to the special $R$-values yielding analytical
solutions.\\

The (unnormalized) wavefunction, radius and energy for the ground state ($m=0$) and $n=1$ are
given by

\begin{equation}
\label{eq:g10}
\Psi_{1,0} (r_{12}) = s_{0} + s_{1} \, r_{12} , \quad R^{2}_{1,0} = \frac{\delta}{4\gamma}, \quad E_{1,0} = \gamma\,  ,
\end{equation}

\noindent
where from now on we have denoted $s_{k,0}\equiv s_{k}$.
The parameters $\delta$ and $\gamma$ are tabulated in Table \ref{tab:table1}.

\begin{table}[H]
  \centering
\begin{tabular}{|c|c|c|c|c|c|}
\hline
State & $(n,m)$   & Configuration &$ \chi({\bf \Omega_1}, {\bf \Omega_2})$    & $\delta$ & $\gamma$ \\ \hline
${}^1 S$ & (1,0) & $s^2$ & $1$ & $2(d-1)-1$ & $\frac{1}{(d-1)-1}$ \\ \hline
\end{tabular}
\caption{Ground state for $n=1$ of $(d-1)$-spherium.}
\label{tab:table1}
\end{table}

\noindent The (unnormalized) ground state wavefunctions for $n=2,3$
are the following,

\begin{eqnarray}
\label{eq:n2}
\Psi_{2,0}(r_{12}) &=& s_{0} + s_{1}\,r_{12}  + s_{2}\,r_{12}^{2} \\
\label{eq:n3}
\Psi_{3,0}(r_{12}) &=& s_{0} + s_{1}\,r_{12}  + s_{2}\,r_{12}^{2} + s_{3}\,r_{12}^{3},
\end{eqnarray}

\noindent
where the coefficients $s_k \equiv s_{k,0}(d)$, obtained through the recurrence
relation (\ref{eq:recuterms}), are analytically given in Appendix A for $k=1,2,3$ and numerically shown in Table \ref{tab:tablecoef} for $d=3-6$.
\begin{table}[H]
  \centering
  \begin{tabular}{|c | c | c | c | c| }
\hline
$d$ & $s_{0}$ & $s_{1}$ & $s_{2}$  & $s_{3}$   \\
 \hline
3  & 1 & 1 & 0.178571 & 0.012946     \\[0.5em]
4  & 1 & 0.5 & 0.053030 & 0.002703 \\[0.5em]
5  & 1 & 0.333333 & 0.025 & 0.000968  \\[0.5em]
6  & 1 & 0.25 & 0.014474 & 0.000449  \\[0.25em]
\hline
\end{tabular}
\caption{Numerical values of the expansion coefficients $s_k \equiv s_{k,0}$ for $d=3,4,5,6$ and $n=1,2,3$.}
\label{tab:tablecoef}
\end{table}


In order to compute the entanglement of the
spherium's eigenstates (with $m=0$) we are
going to work with appropriately normalized
eigenfunctions,

\begin{equation}
\psi_{n,0} = \frac{\Psi_{n,0}}{R^{d-1} N_n^{1/2}},
\end{equation}
where $N_n = \int |\Psi_{n,0}|^2 \, d\Omega_1 d\Omega_2$.
The wavefunctions $\psi_{n,0}$ are now normalized to one over
the surface of a hyper-sphere of radius $R$:
$\int |\psi_{n,0}|^2 \, R^{2(d-1)} d\Omega_1 d\Omega_2 = 1$.
The analytical values of the constant $N_n$ are determined in the
next section.\\


\section{Entanglement in $(d-1)$-spherium}

Let us now evaluate in an analytical way the entanglement for the wavefunctions $\psi_{n,0}(r_{12})$ of the $(d-1)$-spherium as described in the two previous sections. For this purpose we need to calculate first the constant $N_n = \int |\Psi_{n,0}|^2 \, d\Omega_1 d\Omega_2$, and then the trace $Tr\left[\left(\rho_{1}^{(coord.)}\right)^{2}\right]$ which is given by the following multidimensional definite integral

\begin{eqnarray}
\label{eq:3}
Tr\left[\left(\rho_{1}^{(coord.)}\right)^{2}\right] &=&
\int_{\mathbb{R}^{4(d-1)}} \psi_{n,0}(\vr_{1}',\vr_{2}) \psi_{n,0}^{*}(\vr_{1},\vr_{2}) \times \nonumber \\
& & \hspace{1.2cm}\psi_{n,0}^{*}(\vr_{1}',\vr_{2}') \psi_{n,0}(\vr_{1},\vr_{2}')\times \nonumber \\     			 & & \hspace{1.2cm} R^{4(d-1)}    d\Omega_{1}d\Omega_{2}d\Omega_{1}'d\Omega_{2}', \nonumber \\
\end{eqnarray}

\noindent
where  $R^{d-1} d\Omega_k$, $k=1,2$ are area elements
on the surface of a $(d-1)$-hypersphere, and $d\Omega_k$
are elements of hyper-spherical angle, given by

\begin{equation}
d\Omega_k = \left(\prod_{j=1}^{d-2} \sin^{d-j-1}\theta_{j}^{(k)} \right) d\phi^{(k)}.
\end{equation}

\noindent
in terms of the hyperspherical angular coordinates of the two particles $\{ \theta^{(k)}_1, ..., \theta^{(k)}_{d-2}, \phi^{(k)} \}$,
with $0\leq \theta^{(k)}_{j} \leq \pi$ for $j=1,\ldots,d-2$, and $0\leq \phi^{(k)} \leq 2\pi$. Atomic units will be used throughout the rest of the paper.\\

Here a comment concerning coordinates is in order.
In Section II, when discussing general aspects of entanglement,
the integrals involved in the calculation of entanglement
were expressed in cartesian coordinates. However, in the particular
case of spherium, it is clear that the most natural coordinates to
employ are the hyper-spherical ones. Hence, as already indicated by
the elements $d\Omega_i$ appearing in (\ref{eq:3}), in the present work
we are going to formulate all the relevant integrals first
in terms of hyper-spherical coordinates on the $(d-1)$-sphere
where the two electrons are confined. For technical reasons we are also
going to define a new set of angular variables in order to actually
compute the aforementioned integrals. \\

To solve some of the integrals appearing in the study of entanglement
in spherium we
shall apply the methodology recently developed by Ruiz \cite{BR09}
to deal with atomic-related integrals. Let us first calculate the
normalization constant $N_1$ of the ground state wavefunction
$\Psi_{1,0}(r_{12})$ given by eq. (\ref{eq:g10}); that is,

\begin{equation}
\label{eq:norm1}
N_1 = \int |\Psi_{1,0}|^2 \, d\Omega_1 d\Omega_2 = J_0 + 2 \gamma J_1 + \gamma^{2} J_2,
\end{equation}

\noindent
where the symbols $J_k, k = 0,1,2,$ denote the integral functions

\begin{equation}
\label{eq:int22}
J_k \equiv  \int r_{12}^{k}\, d\Omega_{1} d\Omega_{2}, \quad k = 0,1,2 	
\end{equation}

\begin{figure}
\includegraphics[]{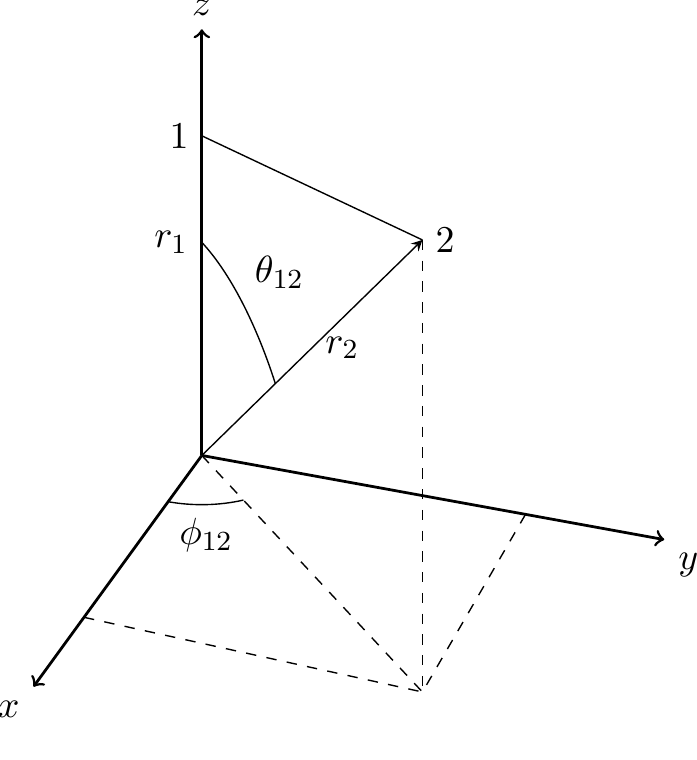}
\caption{Definition of the electron's coordinates used
for the evaluation of entanglement-related integrals} \label{fig:axes}
\end{figure}

To evaluate these integrals we begin by doing a change of variables. Consider the triangle formed by the vectors
$\vr_{1}$,  $\vr_{2}$, and $\vr_{12}$, where the last one stands for the relative vector
position of particle 2 with respect to particle 1 (see Fig. \ref{fig:axes}). Following
an idea originally advanced by Calais and L\"owdin \cite{calais}, we rotate the coordinate frame
used to define the angular spherical coordinates of the vector $\vr_{2}$. The $z$ axis of the
new frame is the line joining the origin (which is the same as in the original frame) with
particle 1, with the positive direction towards particle 1. The angular coordinates of $\vr_{2}$ in the new frame
are now denoted $\{ \theta_{1}^{(12)}, ..., \theta_{d-2}^{(12)}, \phi^{(12)} \}$ (see Figure 1 for
a three dimensional illustration of this change of reference frame). The integration variables
concerning particle 2 are then transformed as: $\theta_i^{(2)} \rightarrow \theta_i^{(12)}$,
and $\phi^{(2)} \rightarrow \phi^{(12)}$. The volume element associated to electron $2$ can
then be re-cast as,

\begin{eqnarray}
\label{}
d\Omega_{2} &=& \left(\prod_{j=1}^{d-2} \sin^{d-j-1}\theta_{j}^{(2)}\right) d\phi^{(2)} \nonumber \\
             &=& \left(\prod_{j=1}^{d-2} \sin^{d-j-1}\theta_{j}^{(12)}\right) d\phi^{(12)} \nonumber \\
             &=& d\Omega_{12}.
\end{eqnarray}

\noindent
Moreover, we use the Cohl representation \cite{cohl} for $r_{12}^{p}$ in terms of the orthogonal Gegenbauer polynomials  $C_{n}^{\alpha}(x)$:

\begin{equation}
\begin{split}
r_{12}^{p} & =  \sum_{n=0}^{\infty} \frac{(d+2 n-2) \, 2^{d+p-3}\, \Gamma \left(\frac{d-2}{2}\right) \left(-\frac{1}{2}\right)_n R^p\, }{\sqrt{\pi }\, \Gamma \left(d+n+\frac{p}{2}-1\right)} \\
& \times \Gamma \left(\frac{1}{2} (d+p-1)\right) C_{n}^{d/2-1}(\cos \theta_{12})\\
 & =  \sum_{n=0}^{\infty} -\frac{\pi ^{\frac{d}{2}-1} 2^{d+p-2} \Gamma \left(n-\frac{1}{2}\right) R^p \Gamma \left(\frac{1}{2} (d+p-1)\right)}{\Gamma \left(d+n+\frac{p}{2}-1\right)} \\
 &   \times \sum_{\{\mu \}} \mathcal{Y}^{*}_{n,\{\mu \}}(\Omega_{2})\mathcal{Y}_{n,\{\mu\}}(\Omega_{1}), \,\, p = 1,2,\ldots,
 \label{eq:rep12}
\end{split}
\end{equation}
where $\Omega_k = (\theta^{(k)}_{1},\theta^{(k)}_{2},\ldots,\theta^{(k)}_{d-2},\theta^{(k)}_{d-1}\equiv\phi^{(k)})$, and $\mathcal{Y}_{n,\{\mu\}}$ denote the known hyperspherical harmonics, which have the following expression \cite{rompabjes, dehros, avery}
\begin{eqnarray}
\label{eq:hypersphhar}
\mathcal{Y}_{l,\{\mu\}}(\Omega_{d-1}) &=& A_{l,\{\mu\}}e^{im\phi}\prod_{j=1}^{d-2}C_{\mu_{j}-\mu_{j+1}}^{\alpha_{j}+\mu_{j+1}}(\cos \theta_{j})\nonumber \\
& & \hspace{2.8cm}\times (\sin \theta_{j})^{\mu_{j+1}}\nonumber \\
 &=& \frac{1}{\sqrt{2\pi}}e^{im\phi} \prod_{j=1}^{d-2}\tilde{C}_{\mu_{j}-\mu_{j+1}}^{\alpha_{j}+\mu_{j+1}}(\cos \theta_{j})\nonumber \\
 & & \hspace{2.5cm}\times (\sin \theta_{j})^{\mu_{j+1}},
\end{eqnarray}
with $\alpha_{j}=(d-j-1)/2$, and the normalization constant is
\begin{eqnarray}
\label{eq:normaconshyperspheriharm}
|A_{l,\{\mu\}}|^{2} &=& \frac{1}{2\pi}\prod_{j=1}^{d-2}\frac{(\alpha_{j}+\mu_{j})(\mu_{j}-\mu_{j+1})![\Gamma(\alpha_{j}+\mu_{j+1})]^{2}}{\pi\, 2^{1-2\alpha_{j}-2\mu_{j+1}}\Gamma(2\alpha_{j}+\mu_{j}+\mu_{j+1})}\nonumber \\
&\equiv & \frac{1}{2\pi} \prod_{j=1}^{d-2} A_{\mu_{j},\mu_{j+1}}^{(j)}\, .
\end{eqnarray}
The symbols $ C_{m}^{\alpha}(x)$ and $ \tilde{C}_{m}^{\alpha}(x)$ denote the orthogonal and orthonormal Gegenbauer polynomials \cite{olver} of degree $m$ and parameter $\alpha$ with respect to the weight function $\omega_{\alpha}^{*} = (1-x^{2})^{\alpha-\frac{1}{2}}$ on the interval $[-1,+1]$, respectively,  so that
\begin{equation}
\label{eq:gegenpoly}
\tilde{C}_{m}^{\alpha}(x)= \left(\frac{m!(m+\alpha)\Gamma^{2}(\alpha)}{\pi\,2^{1-2\alpha}\Gamma(2\alpha+m)}\right)^{\frac{1}{2}}C_{m}^{\alpha}(x).
\end{equation}						

Then, by using the expressions (\ref{eq:int22}) and (\ref{eq:rep12}) as explained in detail in Appendix B, we obtain that the integrals $J_k, k = 0-2,$ are given by
\begin{equation}
\label{eq:J0}
J_0  = \left(\frac{2\pi^{d/2}}{\Gamma(\frac{d}{2})}\right)^{2},
\end{equation}

\begin{equation}
\label{eq:J1}
J_1 = \frac{2^{d+1} \pi ^{d-\frac{1}{2}} }{\Gamma \left(d-\frac{1}{2}\right)}R
\end{equation}
and
\begin{equation}
\label{eq:J2}
J_2 =\frac{8 \pi ^d }{\Gamma \left(\frac{d}{2}\right)^2}R^{2}
\end{equation}
respectively. Finally, these values together with eq. (\ref{eq:norm1}) allows us to write the normalization constant as
\begin{equation}
\label{eq:norm1final}
N_1 =4 \pi ^d \left(\frac{1+ 2 \gamma ^2 R^2}{\Gamma \left(\frac{d}{2}\right)^2}+\frac{2^d \gamma }{\sqrt{\pi }\, \Gamma \left(d-\frac{1}{2}\right)}R\right)
\end{equation}
\noindent
Replacing the analytical expressions for the spherium state $\psi_{1,0}(r_{12})$, into the
general expression for $Tr [(\rho_{1}^{(coord.)})^2]$ one gets,
\begin{eqnarray}
\label{eq:traza1}
&&  Tr[(\rho_{1}^{(coord.)})^{2}] \nonumber \\
&=& R^{2d-2}\int |\rho_{1}^{(coord.)}(\vr_{1}',\vr_{1}) |^{2} \, d\Omega_{1}' d\Omega_{1} \nonumber \\
& =& R^{4d-4} \int \psi_{1,0}(\vr_{1}',\vr_{2})\psi_{1,0}^{*}(\vr_{1},\vr_{2})\psi_{1,0}^{*}(\vr_{1}',\vr_{2}')\times\nonumber \\
& & \psi_{1,0}(\vr_{1},\vr_{2}') \,
d\Omega_{1} d\Omega_{2} \, d\Omega_{1}' d\Omega_{2}' \nonumber \\
&=& N_1^{-2}\, \int_{\mathbb{S}^{d-1}} (1+\gamma\,r_{12}) (1+\gamma\,r_{12'})\times\nonumber \\
& &  (1+\gamma\,r_{1'2}) (1+\gamma\,r_{1'2'})\,
 d\Omega_{1} d\Omega_{2} \, d\Omega_{1}' d\Omega_{2}' \nonumber  \\
 \label{eq:traza11}
&=& N_1^{-2}\,\int \Bigg[ 1 + \gamma (r_{12}+r_{12'}+r_{1'2}+r_{1'2'}) \nonumber \\
&+& \gamma^{2} (r_{12}r_{12'}+r_{12}r_{1'2}+r_{12'}r_{1'2}+r_{12}r_{1'2'}\nonumber \\
&+& r_{12'}r_{1'2'}+r_{1'2}r_{1'2'}) \nonumber  \\
&+&  \gamma^{3} (r_{12}r_{12'}r_{1'2}+r_{12}r_{12'}r_{1'2'}+r_{12}r_{1'2}r_{1'2'}\nonumber\\
&+& r_{12'}r_{1'2}r_{1'2'}) + \gamma^{4}\,r_{12}r_{12'}r_{1'2} r_{1'2'} \Bigg] \nonumber \\
&\times & \, d\Omega_{1} d\Omega_{2} d\Omega_{1}' d\Omega_{2}'
\end{eqnarray}
	
For convenience and taking into account the symmetries of the integrand of (\ref{eq:traza11}), we rewrite this expression as
\begin{equation}
\label{eq:traza1bis}
Tr[(\rho_{1}^{(coord.)})^{2}] = N_1^{-2} (I_{0}+ 4\gamma\,I_{1}+ 6\gamma^{2}\, I_{2}+ 4\gamma^{3}I_{3}+ \gamma^{4}\,I_{4}),
\end{equation}
where the symbols $I_i, i = 1-4,$ denote the following integral functions:

\begin{equation}
\label{eq:I0}
I_0 \equiv \int  d\Omega_{1} d\Omega_{2} d\Omega_{1'} d\Omega_{2'},
\end{equation}

\begin{equation}
\label{eq:I1}
I_1 \equiv \int r_{12}\, d\Omega_{1} d\Omega_{2} d\Omega_{1'} d\Omega_{2'},
\end{equation}

\begin{equation}
\label{eq:I2}
I_2 \equiv \int r_{12} \, r_{12'}\ d\Omega_{1} d\Omega_{2} d\Omega_{1'} d\Omega_{2'},
\end{equation}

\begin{equation}
\label{eq:I3}
I_3 \equiv \int r_{12} \, r_{12'}\,r_{1'2} \, d\Omega_{1} d\Omega_{2} d\Omega_{1'} d\Omega_{2'},
\end{equation}

\begin{equation}
\label{eq:I4}
I_4 \equiv \int r_{12} \, r_{12'}\,r_{1'2} \,r_{1'2'}\, d\Omega_{1} d\Omega_{2} d\Omega_{1'} d\Omega_{2'}.
\end{equation}
These integrals have been analytically evaluated by means of the methodology described in Appendix B, obtaining the following values:
\begin{eqnarray}
\label{eq:i0}
I_{0}  &=& \left(\frac{2\pi^{\frac{d}{2}}}{\Gamma(\frac{d}{2})}\right)^{4}        \\
\label{eq:i1}
I_{1}  &=& \frac{2^{d+3} \pi ^{2 d-\frac{1}{2}} }{\Gamma \left(d-\frac{1}{2}\right) \Gamma \left(\frac{d}{2}\right)^2}R\\
\label{eq:i2}
I_{2}  &=& \frac{4^{d+1} \pi ^{2 d-1} }{\Gamma \left(d-\frac{1}{2}\right)^2}R^2\\
\label{eq:i3}
I_{3}  &=&	 \frac{2^{3 d+1} \pi ^{2 d-\frac{3}{2}}  \Gamma \left(\frac{d}{2}\right)^2}{\Gamma \left(d-\frac{1}{2}\right)^3}R^3\\
\label{eq:i4}
I_{4}  &=&2^{4 d-3} \pi ^{2 d-2} \left[\frac{\Gamma \left(\frac{d}{2}\right)}{\Gamma\left(d+\frac{1}{2}\right)}\right]^{4} R^4\times \nonumber\\
& & \hspace{-1.8cm}\Bigg[  \,_5F_4\left(\frac{1}{2},\frac{1}{2},\frac{1}{2},\frac{1}{2},d-1;d+\frac{1}{2},d+\frac{1}{2},d+\frac{1}{2},d+\frac{1}{2};1\right)\nonumber \\
& &\hspace{-1.6cm}+8\left(d-\frac{1}{2}\right)^{4}\times\nonumber\\ & &
\hspace{-1.8cm} _5F_4\left(-\frac{1}{2},-\frac{1}{2},-\frac{1}{2},-\frac{1}{2},d-2;d-\frac{1}{2},d-\frac{1}{2},d-\frac{1}{2},d-\frac{1}{2};1\right)\Bigg]\nonumber \\
\end{eqnarray}


Then, taking into account  (\ref{eq:e6}) and (\ref{eq:traza1bis}) we obtain that the entanglement measure for the spherium state $\psi_{1,0}(r_{12})$ is given by

\begin{eqnarray}
\label{eq:traza2}
\xi[|\psi_{1,0}\rangle] &=& 1 - Tr[(\rho_{1}^{(coord.)})^{2}] \nonumber \\
						& = & 1- N_1^{-2} (I_{0}+ 4\gamma\,I_{1}+ 6\gamma^{2}\, I_{2}+ 4\gamma^{3}I_{3}\nonumber\\
						& & +\, \gamma^{4}\,I_{4})
\end{eqnarray}

\noindent
The calculations required for evaluating the normalization constants $N_n$ and the integrals $I_i, i =1-4$ involved in the determination of the amount of entanglement of the spherium $s$-eigenstates $\psi_{n,0}(r_{12})$ with $n \ge 2$ are similar to those for the state $\psi_{1,0}(r_{12})$, following the lines indicated in Appendix B. In particular, the above explained analytical techniques can be readily applied to the $n=2$ and $n=3$
$s$-states, with the wavefunctions given by Eqs. (\ref{eq:n2})-(\ref{eq:n3}) and Table \ref{tab:tablecoef}.\\

The results obtained for the amount of entanglement exhibited by the $(d-1)$-spherium (singlet)
ground state are summarized in Table \ref{tab:table2} and in Figures \ref{fig:2}- \ref{fig:4}.
In Table \ref{tab:table2} we provide the amounts of entanglement and the energies corresponding
to the ground state of $(d-1)$-spherium for various dimensionalities. The analytical procedure for
calculating the entanglement of $s$-states of spherium has been checked by the numerical
computation of entanglement for some of these states.

\begin{table}[h]
  \centering
\begin{tabular}{|c|c|c|c|c|}
\hline
State & $d$ & $R_{n,0}$ & $E_{n,0}$ & $\xi[|\Psi_{n,0}\rangle]$\\ \hline
\multirow{3}{*}{ $n=1$} & $3$ &$0.866025$ \ & $1$ & $0.0677386$ \\
 & $4$ & $1.58114$   & $0.5$ & $0.0436006$                  \\
 & $5$ & $2.29129$ &  $0.333333$ & $0.0323117$      \\
 & $6$ & $3.$	&$0.25$ &	$0.0256836$		\\
\hline
\multirow{3}{*}{$n=2$} & $3$ &$2.64575$ \ & $0.285714$ & $0.235892$ \\
 & $4$ & $4.06202$   & $0.181818$ & $0.160622$   \\
 & $5$ & $5.47723$ &  $0.133333$ &  $0.121691$   \\
  & $6$ & $6.89202$&	$0.105263$ &	$0.0979235$				\\
\hline
\multirow{3}{*}{$n=3$} & $3$ &$5.43118$  & $0.127128$ & $0.391247$ \\
 & $4$ & $7.51536$   & $0.0929523$ & $0.293556$                \\
 & $5$ & $9.61594$ &  $0.0729996$ & $0.232591$      \\
  & $6$ & $11.7241$&	$0.0600194$	&	$0.191796$					\\
  \hline
  \end{tabular}
\caption{Radius, energy and entanglement values of the of the $(d-1)$-dimensional spherium with singlet ground-state wavefunctions $\Psi_{n,0}(r_{12})$, $n=1,2,3$, for various dimensionalities $d = 3,4,5,6$.}
\label{tab:table2}
\end{table}

It can be seen in Table \ref{tab:table2} that, for a given dimensionality $d$, the amount of entanglement associated with the
ground state of spherium increases with the radius $R$. This trend is akin with what has been recently observed in other
two-electron models \cite{MPDK10}; see also the recent review \cite{TMB10}. In fact, we know from previous experience with two-electron systems that, for a constant strength of the interaction between the particles,
 entanglement tends to increase when the confinement becomes weaker. This behaviour has been observed in several
 systems, such as the Moshinsky model, the Hooke atom, the Crandall model, and the Helium iso-electronic series \cite{MPDK10}.
 The connection between entanglement and confinement has also been detected in two-electron systems
 in a uniform magnetic field \cite{BMPSD2012}. In these systems confinement decreases, and entanglement increases,
  when the intensity of the applied magnetic field becomes weaker. In spherium confinement decreases, and entanglement increases, for increasing values of the radius $R$.
\begin{figure}
\includegraphics[width=0.5\textwidth]{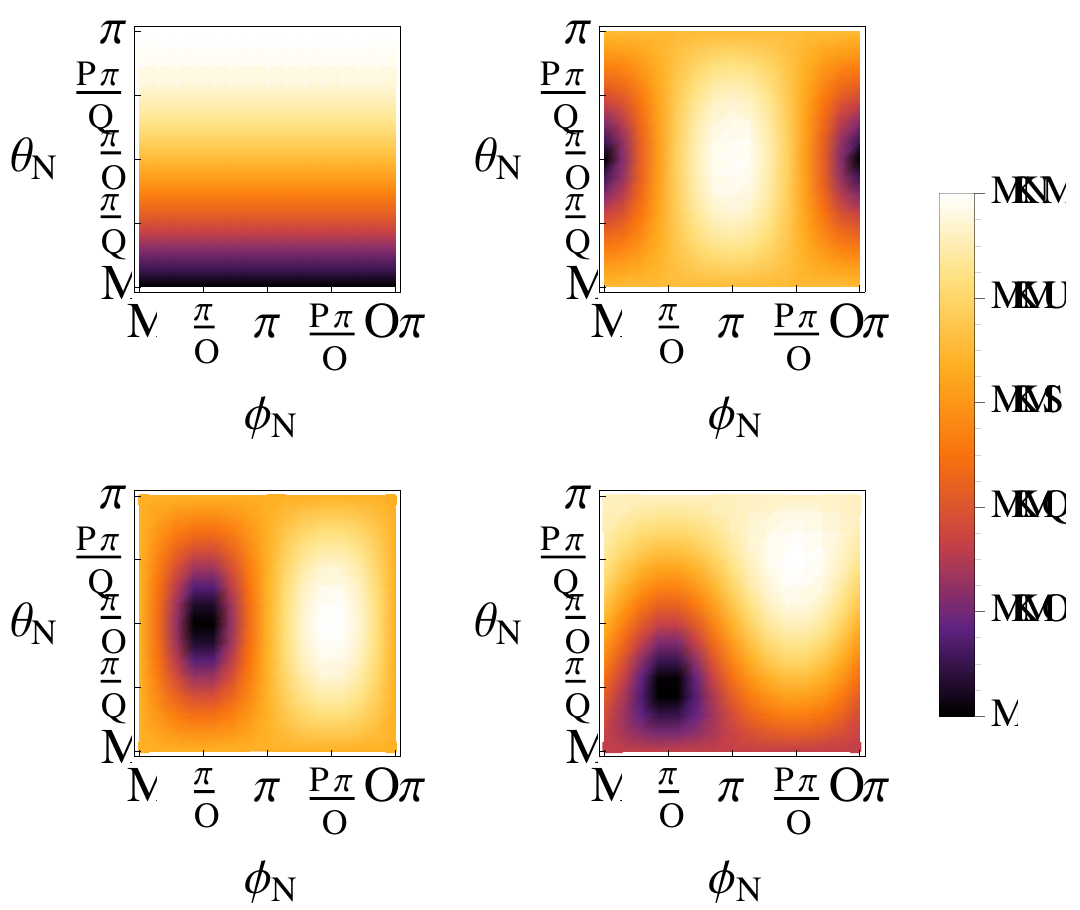}
\caption{Wave function $\psi(\theta_1,\phi_1,\theta_2,\phi_2)$
as a function of the angular coordinates $\theta_1,\phi_1$
of one the electrons for constant values $\theta_2$ and $\phi_2$
of the coordinates of the other electron. On the upper left plot we have
$\theta_2=0$ and $\phi_2=0$ and on the upper right one $\theta_2=\frac{\pi}{2}$ and $\phi_2=0$.
The lower left plot corresponds to
$\theta_2=\frac{\pi}{2}$ and $\phi_2=\frac{\pi}{2}$ and 
the lower right one to $\theta_2=\frac{\pi}{4}$ and $\phi_2=\frac{\pi}{2}$.
The different aspect of the four figures illustrates
the fact that the wavefunction $\psi$ is entangled.} \label{colorin12}
\end{figure}
In Figure \ref{colorin12} we plotted, for $d=3$ (that is,
when the two electrons are confined to an ordinary two-dimensional sphere),
the wavefunction $\psi_{10}(\theta_1, \phi_1, \theta_2, \phi_2)$ as a function
of the angular coordinates $(\theta_1, \phi_1)$ of one of the particles,
keeping constant the values of the coordinates $(\theta_2,\phi_2)$
of the other particle (here we use the standard notation for the
polar and azimuthal coordinates on a two dimensional sphere).
Since the wavefunction is in this case real, we depict in
Fig. \ref{colorin12} the wave function $\psi_{10}$ itself,
not its squared modulus.
In Fig. \ref{colorin12} we have $\theta_2 = 0$;
$\phi_2 = 0$ (upper left) and $\theta_2 = \frac{\pi}{2}$;$\phi_2 = 0$ (upper right);
 $\theta_2 = \frac{\pi}{2}$;
$\phi_2 = \frac{\pi}{2}$ (lower left) and
$\theta_2 = \frac{\pi}{4}$;$\phi_2 =\frac{\pi}{2} $ (lower right).
Figure \ref{colorin12} provides an
illustration of the entangled character of the associated two-electron state.
If it were a non-entangled state, it would be of the form
$\Phi(\theta_1,\phi_1)\Phi(\theta_2,\phi_2) \frac{1}{\sqrt{2}} (\chi_{+-}-\chi_{-+})$.
We would have a factorizable spatial wavefunction (remember that the spatial
parts of the wavefunctions corresponding to the states that we are considering are symmetric)
and a singlet spin wavefunction. With a factorized spatial wavefunction
the four graphics depicted in Figure \ref{colorin12} would be identical.
All of them would correspond to $\Phi(\theta_1,\phi_1)$. 
The differences between the four graphics in Figure \ref{colorin12} 
constitute a concrete pictorial illustration of the entanglement of the concomitant
two-electron state.\\

The connection between entanglement and the radius $R$ of the spherium system can be appreciated in Fig. \ref{fig:2}.
In this Figure we plotted the amount of entanglement versus the radius of the confining sphere for several singlet states wavefunctions $\psi_{n,0}(r_{12})$ of the 2-dimensional ($d=3$) spherium, with 
the integer parameter $n$ (characterizing the radius' values $R_{n,m}$ leading to exact analytical solutions) adopting values $n=1,\ldots,6$. We observe that entanglement grows with the radius. The dependence of entanglement on the spherium radius is, however, nonlinear.
For small values of $R$ (corresponding to small values of the parameter  $n$) the rate of growth of entanglement
with $R$ is greater than for larger values of $R$. The monotonically increasing behaviour of entanglement with $R$ illustrated in Fig. \ref{fig:2} corresponds to the particular case $d=3$. However, the same trend is observed for other values of the spatial dimensionality $d$, as can be seen in Table III for dimensions up to $d=6$.
\begin{figure}[H]
  \centering
    \includegraphics[width=0.43\textwidth]{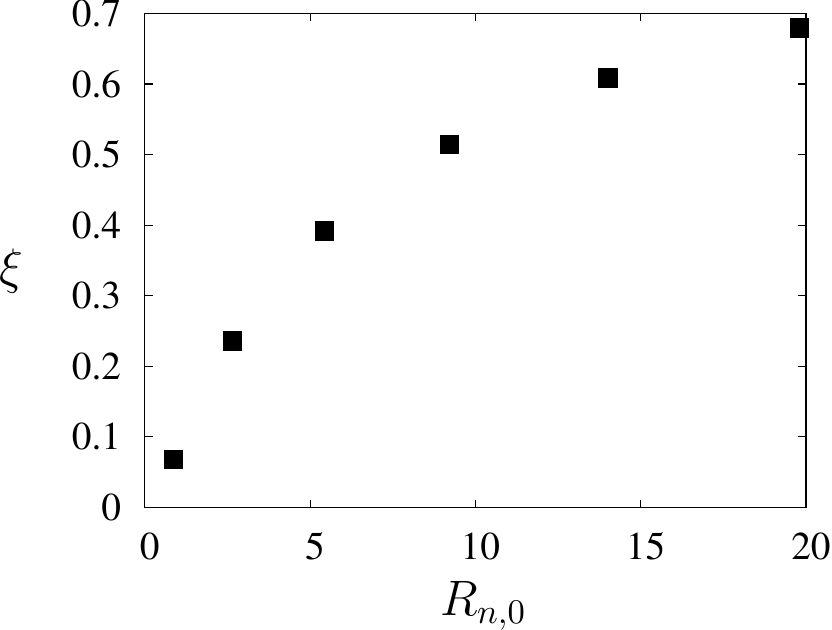}
     \caption{Entanglement against the radius $R$ for the singlet state wavefunctions $\psi_{n,0}(r_{12})$, $n=1,\ldots,6$, of $(d-1)$-dimensional spherium with $d= 3$.}
     \label{fig:2}
\end{figure}
The observed decreasing rate of growth of entanglement with the spherium's radius $R$
does not rule out the possibility that the (increasing) entanglement measure $\xi$ tends to its maximum possible value ($\xi=1$)
in the limit $R\to \infty $. Unfortunately, the case by case (exact) evaluation of the entanglement of each exactly
solvable eigenstate of spherium does not allow us to analytically determine the aforementioned limit value.
However, the behaviour of other two-electron systems suggests that entanglement in spherium does indeed
approach its maximum value as $R\to \infty$. In the $d=3$ case the $R\to \infty $
limit of spherium can be related to a limit case of the two dimensional Hooke atom.
In the $R\to \infty$ limit, as the radius of curvature of the confining sphere tends to zero,
the Schr\"odinger equation describing spherium approaches that of two electrons moving in
a two dimensional Euclidean plane. This suggests that the limit value of entanglement in $d=3$
spherium should coincide with the limit value of entanglement in the two dimensional Hooke system
when the confining potential becomes negligible compared with the electron-electron interaction
potential. Results reported by  Ko\'scik and Hassanabadi in \cite{KH2012}
provide numerical evidence that the entanglement of the two dimensional
Hooke system tends to its maximum value in this limit. \\

 \begin{figure}[H]
  \centering
    \includegraphics[width=0.43\textwidth]{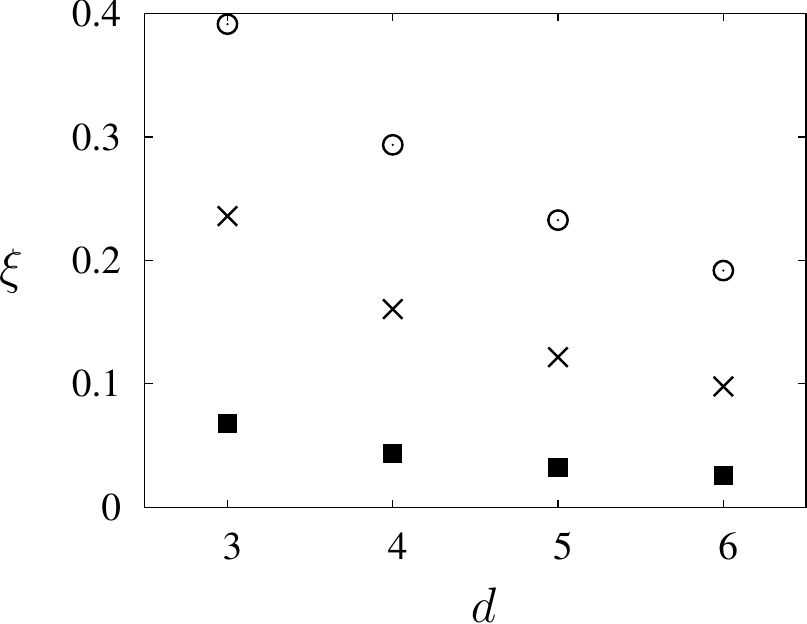}
      \caption{Entanglement against dimensionality for the singlet state wavefunctions $\psi_{n,0}(r_{12})$ with $n=1\,(\blacksquare)$, $n=2\,(\times)$, $n=3\,(\odot)$, of $(d-1)$-dimensional spherium with $d=3,\ldots,6$.}
      \label{fig:3}
\end{figure}

\begin{figure}[H]
  \centering
    \includegraphics[width=0.43\textwidth]{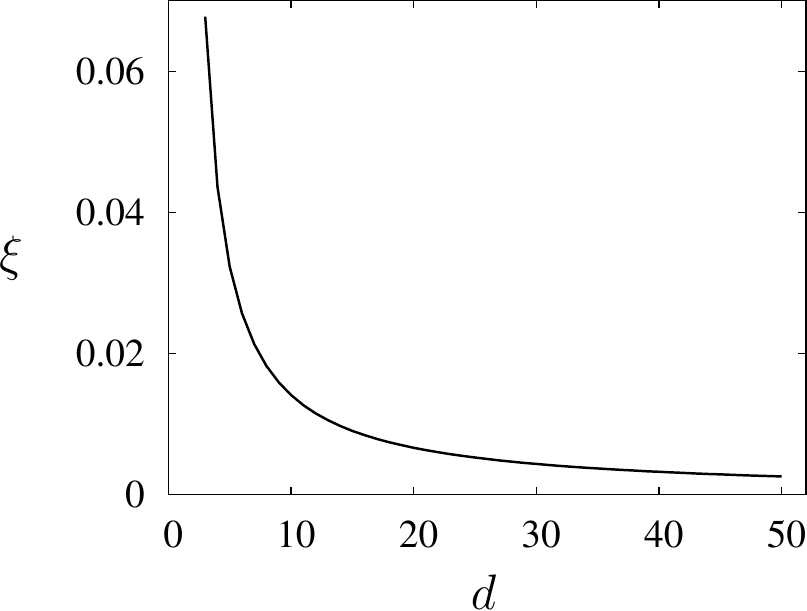}
      \caption{Entanglement against dimensionality for the singlet state wavefunction $\psi_{1,0}(r_{12})$.}
      \label{contindim}
\end{figure}

In Fig. \ref{fig:3} the ground state's entanglement is plotted  against the spatial dimensionality $d$.
We have computed the entanglement measure based on the linear entropy of the single
particle reduced density matrix for spatial dimensionalities in the range $3 \le d \le 6$, and for $n=1,2,3$.
We see that the range of possible values of entanglement, as well as the largest adopted value
(for the above range of $n$-values) decreases with $d$.\\

 It can be appreciated from Fig. \ref{fig:3} that, for given constant values of 
 the integer parameter $n$ determining the special radius $R_{n,m}$ 
 for which spherium admits closed analytical solutions,
  the amount of entanglement exhibited by the ground state of spherium
decreases monotonically with the spatial dimensionality. We conjecture that
entanglement behaves in this way for all values of the parameter $n$.
In the particular case of $n=1$, since we have an analytical expression for
$R_{10}$, we can obtain a closed analytical expression for the
entanglement of the states $\psi_{10}$ for all values of the spatial dimension $d$.
The corresponding behaviour of entanglement as a function of $d$ is shown in Fig.
\ref{contindim},
where it can be seen that entanglement decreases with $d$. This trend might be
related to a well-known, but counterintuitive, feature of multi-dimensional spheres:
the surface area of a $(d-1)$-hypersphere of radius $1$ (that is, the total hyper-solid
angle $\int_{\rm sphere} d\Omega$) tends to zero as $d\to \infty$
(for an interesting discussion on the physical implications of the geometry
of hyperspheres see, for instance, \cite{BT1988} and references therein).
The above can be construed as implying that, as far as the hyperspherical
angular degrees of freedom are concerned, the particles constituting the spherium system can be
regarded as becoming more confined as $d$ increases.  These
geometrical considerations suggest a tentative explanation of
the behaviour of entanglement with spatial dimensionality $d$ in spherium:
entanglement decreases with $d$, because an increasing spatial dimensionality tends to make the system more confined.
Then, according to this explanation, the entanglement-dimensionality relation in spherium would be another
instance of the entanglement-confinement relation observed in several two-electron systems \cite{MPDK10}.
These considerations have some plausibility in connection with the behaviour of entanglement
with dimensionality for large values of $d$. However, a simple, direct connection between entanglement
and the area of the unit hyper-sphere seems unlikely, since for $n=1$ entanglement decreases monotonically with
the spatial dimension for all $d$-values, while the surface area of a unit hyper-sphere
does not behave monotonically with $d$: for moderately small values of $d$ it first
increases with $d$, reaching a maximum for $d \approx 7$, and then decreases monotonically for
all $d$. The decreasing behaviour of entanglement with spatial dimension
in spherium might be related to the properties of other quantum mechanical models where the
limit of high dimensionality leads to classical behaviour \cite{Y82}. \\

The effect of space dimensionality on entanglement has also been studied
in the Hooke atom by Ko\'scik and Hassanabadi \cite{KH2012}. These authors studied the behaviour
of entanglement in the Hooke system for one, two, and three spatial dimensions. The dependence of entanglement
on spatial dimension is not as clear in the Hooke atom as it is in spherium. Indeed, the dependence of
entanglement with dimension in the Hooke system depends on the strength of the electron-electron
interaction (as compared with the strength of the confining potential). This more complicated behaviour
is probably due to the fact that in the Hooke atom the entanglement features
of the system's eigenstates depend on both the radial and angular
behaviours of the concomitant wavefunctions. In spherium, in contrast,
the effective configuration space solely involves the angular variables. \\

\begin{figure}[H]
  \centering
    \includegraphics[width=0.43\textwidth]{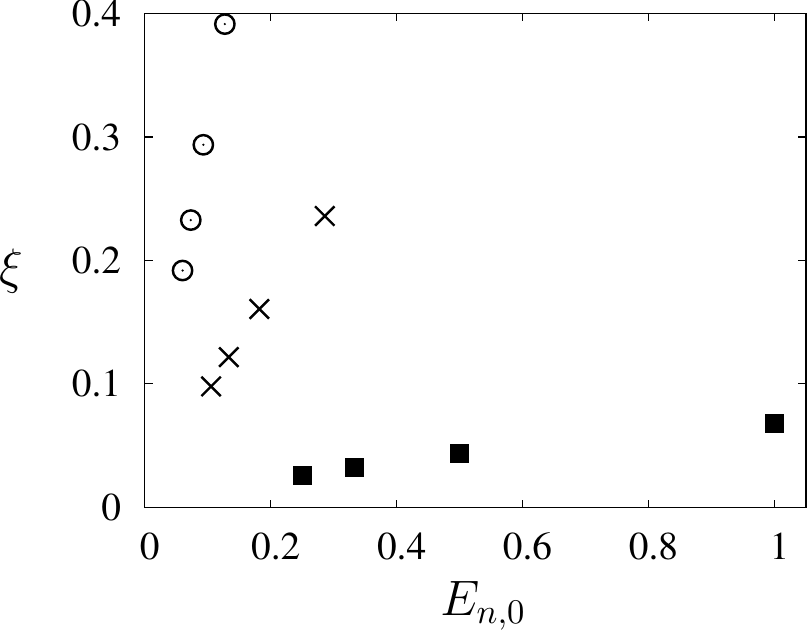}
     \caption{Entanglement against energy for the singlet state wavefunctions $\psi_{n,0}(r_{12})$ with $n=1\,(\blacksquare)$, $n=2\, (\times)$, $n=3\,(\odot)$ of the $(d-1)$-dimensional spherium with $d=3-6$.}
      \label{fig:4}
\end{figure}


In Fig. \ref{fig:4} we depict the amount of entanglement against the energy of the singlet state for $d=3,4,5$ and $6$.
We observe that entanglement of $(d-1)$-dimensional spherium tends to increase with energy.
A similar behaviour has been observed in other models,
 such as the Crandall and the Hooke ones \cite{MPDK10}, as well as for the singlet states of the Helium atom employing
 high-quality, state-of- the-art wavefunctions \cite{DKYPE2012}  (although for more general states of
 Helium the energy-entanglement connection seems to be much more complicated \cite{RS14,RS2015}). \\

Finally, let us comment on the excited states of the spherium. There are no excited states for the singlet wavefunction with $n=1$ and $n=2$. This is due to the fact that the equation allowing for the calculation of the energy for each $n$ has only one root which correspond to $m=0$.
For $n\geq 3$ the excited states begin to appear because the equation mentioned above has a degree equal or greater than $2$.
So e.g., for $n=3$ there are two possible values for $m=0,1$, and therefore one exact analytical excited state
can be obtained (corresponding to $m=1$).


\section{Conclusions}

   We have explored the entanglement related features of $(d-1)$-spherium.
   This quantum system consists of two electrons interacting via a Coulomb
   potential and confined to the surface of a $d$-dimensional ball (that is,
   a $(d-1)$-hypersphere) of radius $R$. This system is quasi-exactly solvable:
   its Schr\"odinger eigenvalue equation can be solved in a closed analytical
   fashion for particular values of the radius $R$ and particular eigenstates.
   In the present contribution we computed in an exact analytical way the amount
   of entanglement (as measured by the linear entropy of the single-particle
   reduced density matrix) of the ground state of spherium, for several values
   of the radius $R$ (corresponding to different values of the parameter $n$)
   and of the space dimension $d$. To the best of our knowledge this is the first two-electron
   system with Coulomb interaction for which exact entanglement calculations
   have been done. We investigated the dependence of entanglement on the
   radius $R$ of the spherium system and on the spatial dimensionality $d$. The
   relation between entanglement and energy was also considered. \\

   We have found that the amount of entanglement of the ground state of spherium increases with the
   radius $R$ of the hypersphere where the particles are confined. This behaviour
   is consistent with a general property exhibited by other two-electron systems:
   entanglement tends to increase when, for a given value of the interaction strength,
   the confinement due to the external common fields acting on both
   particles decreases. For instance, in the helium isoelectronic series the entanglement of
   the ground state increases when one considers decreasing values of the nuclear charge $Z$
   \cite{MPDK10,DKYPE2012}. Likewise, in the three-dimensional Moshinsky model with a uniform magnetic field
   the entanglement of the ground state increases for decreasing values of the applied
   magnetic field \cite{BMPSD2012}.\\

   The results reported in the present work indicate that in spherium the amount of entanglement exhibited by the
   ground state decreases with the spatial dimensionality $d$, a behaviour that can also
   be related to the entanglement-confinement connection. In addition, we have observed
   that entanglement of spherium tends to increase with energy. This
   relation between entanglement and energy is similar to what is observed in other
   two-electron models, such as the Moshinsky system, the Hooke atom, and the
   Crandall model \cite{MPDK10}. \\

   We hope that the techniques developed in the present
   work may stimulate new analytical approaches to the study of
   entanglement in systems with Coulomb interactions. Any further
   developments in this direction will be very welcome.\\

\section*{Acknowledgments}
 This work was partially supported by the Projects
FQM-7276 and FQM-207 of the Junta de Andaluc\'ia and the grant
FIS2011-24540 of the Ministerio de Econom\'ia y Competitividad (Spain).

\appendix

\section{$(d-1)$-Spherium Hamiltonian}

 Here we are going to briefly review some basic aspects of the
 Hamiltonian describing the two-electron system spherium.
 For more details on spherium
 and the solutions of the concomitant Schr\"odinger equation
 see \cite{LG09,LG10} and references therein.
 Spherium consists of two electrons confined to a $(d-1)$-sphere
 (that is, the surface of a $d$-dimensional ball) and interacting
 via a Coulomb potential. The corresponding Schr\"odinger eigenvalue
 equation reads,

 \begin{equation} \label{chesferri}
 -\left(\frac{\hbar^2}{2m} \right) \frac{1}{R^2}
 \left[\Delta^{(1)}_{S^{d-1}} + \Delta^{(2)}_{S^{d-1}} \right] \Psi
 + \frac{e^2}{r_{12}} \Psi = E \Psi,
 \end{equation}

  \noindent
  where $R$ is the radius of the $(d-1)$-sphere, $r_{12}$
  is the distance between the two electrons (evaluated in the $d$-dimensional
  euclidean space in which the $(d-1)$-sphere is embedded), $\Psi$ is
  the two-electron eigenfunction,  $E$ is the corresponding eigenenergy,
  and $\Delta^{(1,2)}_{S^{d-1}} $ are the angular Laplacian operators acting on the
  angular coordinates of each electron. Note that the wavefunction $\Psi$ is a function
  of the hyper-spherical angular coordinates of both electrons,
  $\{ \theta^{(k)}_1, ..., \theta^{(k)}_{d-2}, \phi^{(k)} \}$,
  with $0\leq \theta^{(k)}_{j} \leq \pi$ for $j=1,\ldots,d-2$, and $0\leq \phi^{(k)} \leq 2\pi$.
  The upper index $k=1,2$ refers to the two electrons.
  The spherical Lapacian operator (Laplace-Beltrami operator on the
  $(d-1)$-sphere) acts on a function $f$ defined on the $(d-1)$-sphere
  according to the following recurrence relation,

  \begin{eqnarray} \label{laplasfer}
  \Delta_{S^{d-1}} f(\theta_1, \zeta)   &=&
  \left(\sin \theta_1 \right)^{2-d}  \frac{\partial}{\partial \theta_1}
  \left[\left(\sin \theta_1 \right)^{d-2}
  \frac{\partial f}{\partial \theta_1}  \right] \nonumber \\
  &+& \left(\sin \theta_1 \right)^{-2}
  \Delta_{\zeta} f,
  \end{eqnarray}

  \noindent
  where $\zeta = \{\theta_2, ..., \theta_{d-2},\phi \}$ denotes
  the set of all the angular coordinates on the $(d-1)$-sphere
  except $\theta_1$, and $\Delta_{\zeta} $ is the spherical
  Laplacian corresponding to a $(d-2)$-sphere with hyper-spherical
  coordinates $\zeta = \{\theta_2, ..., \theta_{d-2},\phi \}$.
  That is, the operator $\Delta_{\zeta} $ only involves derivatives
  with respect to the coordinates appearing in the set $\zeta$.
  For instance, for $d=3$ the spherical Laplacian adopts the well known form,

  \begin{equation} \label{laplasferdos}
  \Delta_{S^2} f(\theta, \phi)   =
  \frac{1}{\sin \theta} \frac{\partial}{\partial \theta}
  \left(\sin \theta \,
  \frac{\partial f}{\partial \theta }  \right)
  +  \frac{1}{\sin^2 \theta} \frac{\partial^2 f}{\partial \phi^2}.
  \end{equation}

  It is convenient to recast the Schr\"odinger equation (\ref{chesferri})
  in a dimensionless form, using atomic units. In order to do that
  we divide equation (\ref{chesferri}) by the constant $m e^4/\hbar^2$
  (which has dimensions of energy) obtaining,

  \begin{equation} \label{formitomic}
 -\frac{1}{2 {\tilde R}^2}
 \left[\Delta^{(1)}_{S^{d-1}} + \Delta^{(2)}_{S^{d-1}} \right] \Psi
 + \frac{1}{{\tilde r}_{12}} \Psi = {\tilde E} \Psi,
 \end{equation}

 \noindent
 where,

 \begin{eqnarray}
  {\tilde R} &=&  \frac{m e^2}{\hbar^2} \, R  \label{paramito}  \\
  {\tilde r}_{12} &=& \frac{m e^2}{\hbar^2} \, r_{12}  \\
  {\tilde E} &=&  \frac{\hbar^2}{m e^4} \, E.
  \end{eqnarray}

  \noindent
  Note that in its dimensionless form (\ref{formitomic})
  the Schr\"odinger equation for spherium only has one parameter,
  the dimensionless radius ${\tilde R}$ given by (\ref{paramito}).
  This parameter is a dimensionless quantity involving the parameter
  $e^2$, measuring the strength of the interaction between the electrons, and
  the radius $R$ of the confining sphere. When studying entanglement in spherium
  we investigate, among other things, its dependence on the dimensionless parameter
  ${\tilde R}$, which can be regarded as proportional to the quotient between
  the quantities $e^2$ (interaction strength) and $1/R$ (amount of confinement).
  In the rest of the present article, since we are going to deal exclusively
  with the dimensionless form (\ref{formitomic}) of the spherium's Schr\"odinger
  equation, we are going to drop the upper ``tilde" from $R$, $r_{12}$ and $E$
  (as in equation (\ref{hamilto})).\\

  In the case of $s$-states, the solutions of spherium's Schr\"odinger equation are
  functions of the inter-particle distance $r_{12}$. That is, one has $\Psi = \Psi(r_{12})$,
  with

  \begin{equation}
  r_{12} = R \sqrt{2(1-\cos \alpha)},
  \end{equation}

  \noindent
  where $\cos \alpha$ can be expressed in terms of the hyper-spherical
  coordinates of the two electrons,

  \begin{eqnarray}
  \cos \alpha &=& \cos \theta_1^{(1)} \cos \theta_1^{(2)} \nonumber \\
              &+& \sin \theta_1^{(1)} \sin \theta_1^{(2)}
              \cos \theta_2^{(1)} \cos \theta_2^{(2)} \nonumber \\
               &+& \sin \theta_1^{(1)} \sin \theta_1^{(2)}
              \sin \theta_2^{(1)} \sin \theta_2^{(2)} \nonumber \\
               \cdots \nonumber \\
            &+& \sin \theta_1^{(1)} \sin \theta_1^{(2)} \cdots
              \sin \theta_{d-2}^{(1)} \sin \theta_{d-2}^{(2)}
              \cos \phi^{(1)} \cos \phi^{(2)}. \nonumber \\
          &+& \sin \theta_1^{(1)} \sin \theta_1^{(2)} \cdots
              \sin \theta_{d-2}^{(1)} \sin \theta_{d-2}^{(2)}
              \sin \phi^{(1)} \sin \phi^{(2)}. \nonumber \\
              \end{eqnarray}

\noindent
For $s$-states the Schr\"odinger equation (\ref{chesferri}) can be
re-expressed in terms of the derivatives of the wavefunction
with respect to the variable $u=r_{12}$ \cite{LG09,LG10},

\begin{equation}
\label{schroeII}
\left[\frac{u^2}{4R^2} -1\right] \frac{d^2\Psi}{du^2} +
\left[\frac{u(2d-3)}{4R^2} - \frac{d}{u}  \right] \frac{d\Psi}{du}
+\frac{\Psi}{u} = E \Psi.
\end{equation}

\section{Expansion coefficients of spherium $s$-eigenstates $\Psi_{n,0}(r_{12})$ with $n =1,2,3$}

The analytical expression for the expansion coefficients $s_k \equiv s_{k,0}(d)$ of the spherium $s$-eigenstates $\Psi_{n,0}(r_{12})$ with $n =1,2,3$, have the following form:
{\small
\begin{eqnarray}
\label{eq:coeffs}
s_{0} &=& 1   \nonumber\\
s_{1} &=& \frac{1}{d-2} \equiv \gamma		\nonumber	\\
s_{2} &=&  \frac{1-2 d}{-8 d^3+34 d^2-46 d+20}     	\nonumber \\
s_{3} &=& \frac{1}{48600}\Bigg(-\frac{5520}{(d-2)^2}-\frac{5400}{d-1}+\frac{1600}{1-2 d}+\frac{2393}{d-2}\nonumber \\
& & -\frac{4050}{d^2}+\frac{24975}{d}-\frac{42336}{2 d+1}\nonumber \\
& & + \frac{900 (d (14 d-23)+6) \sqrt{\frac{d (d (64 (d-2) d+169)-78)+9}{d^2 (d (3-2 d)+2)^2}}}{(d-2) (d-1) d (2 d-1)}\Bigg).\nonumber \\
\end{eqnarray}
}
They have been obtained from the recurrence relation (\ref{eq:recuterms}).\\

\section{Evaluation of the multidimensional integrals involved in the entanglement of spherium}
\label{sec:evaluation}

Here we first give some further details of the calculation of the relevant integrals involved in the determination of the normalization constant $N_1$ of the ground state wavefunction $\Psi_{1,0}(r_{12})$ given by Eq. (\ref{eq:g10}). Later on, we provide with further information of the derivation of the integral functions involved in the determination of the entanglement measure of such a state. Finally, we give indications for the similar calculation of the normalization constant $N_n$ and the entanglement measure of the general wavefunctions $\Psi_{n,0}(r_{12})$, $n \ge 2$, of the $d$-dimensional spherium. \\

\textit{Derivation of the normalization constant $N_1$ given by Eq. (\ref{eq:norm1final})}. This issue reduces to prove that the two-center integrals $J_i, i = 0-2,$ defined by expressions (\ref{eq:int22}) have the values given by Eqs. (\ref{eq:J0}), (\ref{eq:J1}) and (\ref{eq:J2}), respectively.
The value (\ref{eq:J0}) of the integral $J_{0}$ is straightforward since it is the product of the volumes of the hyperspheres for each electron. To obtain the values (\ref{eq:J1}) and (\ref{eq:J2}) of $J_{1}$ and $J_{2}$, respectively, we use the Cohl expansion (\ref{eq:rep12}) for $r_{12}^{p}$ in terms of the Gegenbauer polynomials, $C_{m}^{\alpha}(x)$, and then we apply the orthogonality property of these polynomials which reads \cite{olver} as

\begin{equation}
\label{eq:orthoggegen}
\int_{-1}^{1}(1-t^{2})^{\alpha-\frac{1}{2}} C_{n}^{\alpha}(x)C_{m}^{\alpha}(x)\, dx = \frac{\pi\,2^{1-2\lambda}\Gamma(n+2\lambda)}{n!(n+\lambda)[\Gamma(\lambda)]^{2}}\delta_{m,n}
\end{equation}

\noindent
where $Re(\lambda)>-1/2$ with $\lambda\neq 0$ and that $C_{0}^{\alpha}(x) = 1$. Then, we obtain for the integral $J_{1}$ the following expression:

\begin{equation}
\begin{split}
\label{eq:intr12}	
J_{1} &= \sum_{n=0}^{\infty} (2n +d -2) \, R \, \left(-\frac{1}{2} \right)_{n} 2^{d-2}\, \pi^{-1/2} \nonumber \\
& \frac{\Gamma(\frac{d}{2}-1)\Gamma(\frac{d}{2})}{\Gamma(d+n-\frac{1}{2})} \int C_{n}^{\frac{d}{2}-1} (\cos \theta_{12}) \, d\Omega_{1} d\Omega_{2} \nonumber \\
	&= \int d\Omega_{1} \sum_{n=0}^{\infty} (2n +d -2) \, R \, \left(-\frac{1}{2} \right)_{n} 2^{d-2}\nonumber \\
	&  \pi^{-1/2} \frac{\Gamma(\frac{d}{2}-1)\Gamma(\frac{d}{2})}{\Gamma(d+n-\frac{1}{2})}
 \int C_{n}^{\frac{d}{2}-1} (\cos \theta_{12}) \, d\Omega_{12} \nonumber \\
	&= \frac{2\pi^{\frac{d}{2}}}{\Gamma(\frac{d}{2})} \sum_{n=0}^{\infty} (2n +d -2) \, R \, \left(-\frac{1}{2} \right)_{n} 2^{d-2}\nonumber \\
	& \pi^{-1/2} \frac{\Gamma(\frac{d}{2}-1)\Gamma(\frac{d}{2})}{\Gamma(d+n-\frac{1}{2})} \nonumber \\
	&\times  \int C_{n}^{\frac{d}{2}-1} (\cos \theta_{12})\,C_{0}^{\frac{d}{2}-1} (\cos \theta_{12})\nonumber \\
	& \hspace{1.8cm}\times [\sin \theta_{j=1}^{(12)}]^{d-2} \, d\Omega_{j=1}^{(12)} \nonumber  \\
	 &\times \prod_{j=2}^{d-2} \int_{0}^{\pi} [\sin \theta_{j}^{(12)}]^{d-j-1} \, d\theta_{j}^{(12)} \int_{0}^{2\pi} d\phi^{(12)}\\
\end{split}
\end{equation}

\noindent
Now we do change of integration variables,

\begin{eqnarray}
t &=& \cos \theta_{j=1}^{(12)}, \\
dt &=& - \sin \theta_{j=1}^{(12)}\, d\theta_{j=1}^{(12)} \\
&=& - (1-t^{2})^{1/2} \, d\theta_{j=1}^{(12)}, \nonumber
\end{eqnarray}

\noindent
obtaining the following expression

\begin{eqnarray}
J_{1}
&=& \frac{2\pi^{\frac{d}{2}}}{\Gamma(\frac{d}{2})} \sum_{n=0}^{\infty} (2n +d -2) \, R \, \left(-\frac{1}{2} \right)_{n} 2^{d-2}\nonumber \\
& & \pi^{-1/2} \frac{\Gamma(\frac{d}{2}-1)\Gamma(\frac{d}{2})}{\Gamma(d+n-\frac{1}{2})} \nonumber \\
	&\times & \int C_{n}^{\frac{d}{2}-1} (t)\,C_{0}^{\frac{d}{2}-1} (t)(1-t^{2})^{(\frac{d}{2}-1)-\frac{1}{2}} \, dt \nonumber \\ & &  (2\pi) \prod_{j=2}^{d-2} \sqrt{\pi} \frac{\Gamma(\frac{d-j}{2})}{\Gamma(\frac{d-j+1}{2})} \nonumber \\
&=& \frac{2\pi^{\frac{d}{2}}}{\Gamma(\frac{d}{2})} \sum_{n=0}^{\infty} (2n +d -2) \, R \, \left(-\frac{1}{2} \right)_{n} 2^{d-2}\nonumber \\
& & \pi^{-1/2} \frac{\Gamma(\frac{d}{2}-1)\Gamma(\frac{d}{2})}{\Gamma(d+n-\frac{1}{2})}(2\pi) \delta_{n,0}  \nonumber \\	
 &\times & \prod_{j=2}^{d-2} \sqrt{\pi} \frac{\Gamma(\frac{d-j}{2})}{\Gamma(\frac{d-j+1}{2})}\frac{\pi\, 2^{3-d}\Gamma(n+d-2)}{n! (n+\frac{d}{2}-1)\Gamma(\frac{d}{2}-1)^{2}}\nonumber \\
&=& \frac{2^{d+1} \pi ^{\frac{d}{2}+1} \Gamma \left(\frac{d-1}{2}\right)}{\Gamma \left(d-\frac{1}{2}\right)}R\nonumber \\
 & & \times \prod_{j=2}^{d-2} \sqrt{\pi} \frac{\Gamma(\frac{d-j}{2})}{\Gamma(\frac{d-j+1}{2})}\nonumber\\
 & = & \frac{2^{d+1} \pi ^{d-\frac{1}{2}} }{\Gamma \left(d-\frac{1}{2}\right)}R
\end{eqnarray}			

\noindent
for the integral $J_{1}$ which is equal to the wanted value (\ref{eq:J1}). Operating in a similar way we obtain

\begin{eqnarray}
\label{}	
J_{2} &=& \int d\Omega_{1} \int r_{12}^{2} \, d\Omega_{12} \nonumber \\
 &=& \frac{2\pi^{\frac{d}{2}}}{\Gamma(\frac{d}{2})}  \int \Bigg(2\, R^{2} \,  C_{0}^{\frac{d}{2}-1} (\cos \theta_{12})\nonumber \\
 & & - \frac{2\, R^{2}}{d-2}  C_{1}^{\frac{d}{2}-1} (\cos \theta_{12})\Bigg) \, d\Omega_{12}  \nonumber  \\
 &=& \frac{8 \pi ^{\frac{d+3}{2}}  \Gamma \left(\frac{d-1}{2}\right)}{\Gamma \left(\frac{d}{2}\right)^2}R^2\nonumber\\
  & & \times \prod_{j=2}^{d-2} \sqrt{\pi} \frac{\Gamma(\frac{d-j}{2})}{\Gamma(\frac{d-j+1}{2})}\nonumber\\
  &=& \frac{8 \pi ^d }{\Gamma \left(\frac{d}{2}\right)^2}R^2
\end{eqnarray}								

\noindent
which is equal to the wanted value (\ref{eq:J2}). Finally, by performing the sum in (\ref{eq:norm1}) with $J_{0}, J_{1}$ and $J_{2}$, we arrive at final expression (\ref{eq:norm1final}) for the normalization, $N_{1}$.
\\
It is worth to say that it is possible to compute the normalizaton of an arbitrary $s$-state of the $d$-dimensional spherium by means of the integral $J_{0}$ and the general one-center integral

\begin{equation}
\label{eq:normonecenint}
\int r_{12}^{q} \, d\Omega_{1}d\Omega_{2} = \frac{ 2^{d+q}\pi ^{d-\frac{1}{2}}  \Gamma \left(\frac{d+q-1}{2} \right)}{\Gamma \left(\frac{d}{2}\right) \Gamma \left(d+\frac{q}{2}-1\right)}R^q
\end{equation}
with $q\geq 1$.\vspace{-1cm}\\
\\
\\

\noindent
\textit{Derivation of the values (\ref{eq:i0})-(\ref{eq:i4}) for the multicenter integrals $I_i, i = 0-4$ defined by Eqs. (\ref{eq:I0})-(\ref{eq:I4}), which are involved in the entanglement of the $s$-states of the $d$-dimensional spherium}. These values are characterized by the parameter $n$, which determines the radius $R_n$ of the sphere on which the particles are confined. For this issue we have first determined the following general expressions
{\footnotesize
\begin{eqnarray}
\label{eq:q1-q4}
\int r_{12}^{q_{1}}r_{12'}^{q_{2}}r_{1'2}^{q_{3}}r_{1'2'}^{q_{4}}\, d\Omega_{1}d\Omega_{2}d\Omega_{1'}d\Omega_{2'} & =& \pi ^{2 d-2} 2^{4 d+q_{1}+q_{2}+q_{3}+q_{4}-7} \nonumber\\
& &\hspace{-5.5cm}\times \frac{\Gamma \left(\frac{d+q_{1}-1}{2}\right) \Gamma \left(\frac{d+q_{2}-1}{2}\right)}{\Gamma (d-1) \Gamma \left(d+\frac{q_{1}}{2}-1\right) \Gamma \left(d+\frac{q_{1}}{2}\right) \Gamma \left(d+\frac{q_{2}}{2}-1\right) \Gamma \left(d+\frac{q_{2}}{2}\right)} \nonumber\\
& &\hspace{-5.5cm}\times \frac{ \Gamma \left(\frac{1}{2} (d+q_{3}-1)\right) \Gamma \left(\frac{1}{2} (d+q_{4}-1)\right)}{\Gamma \left(d+\frac{q_{3}}{2}-1\right) \Gamma \left(d+\frac{q_{3}}{2}\right) \Gamma \left(d+\frac{q_{4}}{2}-1\right) \Gamma \left(d+\frac{q_{4}}{2}\right)}R^{q_{1}+q_{2}+q_{3}+q_{4}} \nonumber\\
& &\hspace{-5.5cm}\times \Bigg[ 8 \Gamma (d-1) \Gamma \left(d+\frac{q_{1}}{2}\right) \Gamma \left(d+\frac{q_{2}}{2}\right) \Gamma \left(d+\frac{q_{3}}{2}\right) \Gamma \left(d+\frac{q_{4}}{2}\right) \nonumber\\
& &\hspace{-5.5cm} _5F_4\Big(-\frac{1}{2},-\frac{1}{2},-\frac{1}{2},-\frac{1}{2},d-2;d+\frac{q_{1}}{2}-1,d+\frac{q_{2}}{2}-1,\nonumber\\
& & d+\frac{q_{3}}{2}-1,d+\frac{q_{4}}{2}-1;1\Big) \nonumber\\
& &\hspace{-5.5cm} +\Gamma (d-1) \Gamma \left(d+\frac{q_{1}}{2}-1\right) \Gamma \left(d+\frac{q_{2}}{2}-1\right) \Gamma \left(d+\frac{q_{3}}{2}-1\right) \Gamma \left(d+\frac{q_{4}}{2}-1\right)\nonumber\\
& &\hspace{-5.5cm} _5F_4\left(\frac{1}{2},\frac{1}{2},\frac{1}{2},\frac{1}{2},d-1;d+\frac{q_{1}}{2},d+\frac{q_{2}}{2},d+\frac{q_{3}}{2},d+\frac{q_{4}}{2};1\right)\Bigg]
\end{eqnarray}}
for four-center integrals,
\begin{equation}
\begin{split}
\label{eq:q1-q4B}
\int r_{ij}^{q_{1}}r_{ik'}^{q_{2}}r_{pk}^{q_{3}}\, d\Omega_{1}d\Omega_{2}d\Omega_{1'}d\Omega_{2'} &=\\
&\hspace{-4cm} 2^{3d-3+q_{1}+q_{2}+q_{3}}\pi^{\frac{3}{2}(d-1)}R^{q_{1}+q_{2}+q_{3}}\\
&\hspace{-4cm} \frac{\Gamma(\frac{d+q_{1}-1}{2})
\Gamma(\frac{d+q_{2}-1}{2})\Gamma(\frac{d+q_{3}-1}{2})}{\Gamma(d-1+\frac{q_{1}}{2})
\Gamma(d-1+\frac{q_{2}}{2})\Gamma(d-1+\frac{q_{3}}{2})}\frac{2\pi^{d/2}}{\Gamma(d/2)}.
\end{split}
\end{equation}
for three-center integrals,
\begin{equation}
\begin{split}
\label{eq:q1-q4C}
\int r_{ij}^{q_{1}}r_{pk'}^{q_{2}} \, d\Omega_{1}d\Omega_{2}d\Omega_{1'}d\Omega_{2'} &=\\
&\hspace{-4cm} 2^{2d-2+q_{1}+q_{2}}\pi^{d-1}R^{q_{1}+q_{2}}\\
& \hspace{-4cm} \frac{\Gamma(\frac{d+q_{1}-1}{2})
\Gamma(\frac{d+q_{2}-1}{2})}{\Gamma(d-1+\frac{q_{1}}{2})\Gamma(d-1+\frac{q_{2}}{2})}\left(\frac{2\pi^{d/2}}{\Gamma(d/2)}\right)^{2}.
\end{split}
\end{equation}
for two-center integrals, and
\begin{equation}
\begin{split}
\label{eq:q1-q4D}
\int r_{ij}^{q_{1}} \, d\Omega_{1}d\Omega_{2}d\Omega_{1'}d\Omega_{2'} = \frac{2^{d-1+q_{1}}\pi^{\frac{d-1}{2}}R^{q_{1}}\Gamma(\frac{d+q_{1}-1}{2})}{\Gamma(d-1+\frac{q_{1}}{2})}& \\
\times\left(\frac{2\pi^{d/2}}{\Gamma(d/2)}\right)^{3}. \hspace{2cm}&
\end{split}
\end{equation}
for one-center integrals, where the parameters $q_{i} \geq 1$ for $i=1,\ldots, 4$.\\
Apart from the value (\ref{eq:i0}) of the integral $I_0$, which is straightforward, these general multicenter integral expressions allow us to calculate not only the values (\ref{eq:i0})-(\ref{eq:i4}) of the integrals $I_i, i = 1-4$ needed for the entanglement of the ground-state wavefunctions $\Psi_{1,0}(r_{12})$, but also the corresponding integrals involved in the entanglement of the ground-state wavefunctions $\Psi_{n,0}(r_{12})$ with $n \ge 2$, of the $d$-dimensional spherium in an analytical way.


\newpage

\begin{thebibliography}{}

\bibitem{LG09} P. F. Loos and P. M. W. Gill,
Phys. Rev. Lett. {\bf 103}, 123008 (2009).

\bibitem{LG10} P. F. Loos and P. M. W. Gill,
Mol. Phys. {\bf 108}, 10 (2010).


\bibitem{YPD10} R. J. Ya\~nez, A. R. Plastino, J. S. Dehesa, {\it Eur. Phys. J. D} {\bf 56}, 141 (2010).

\bibitem{BMPSD2012}
P. A. Bouvrie, A. P. Majtey, A. R. Plastino, P. S\'anchez-Moreno, J. S. Dehesa,
\newblock Eur. Phys. J. D {\bf 66}, 1 (2012).


\bibitem{MPDK10}
D. Manzano, A. R. Plastino, J. S. Dehesa, T. Koga,
\newblock J. Phys. A: Math. Theor. {\bf 43}, 275301 (2010).







\bibitem{BZ06} I. Bengtsson, K. Zyczkowski,
{\em Geometry of Quantum States: An Introduction
to Quantum Entanglement} (Cambridge:
Cambridge University Press, 2006).

\bibitem{AFOV08} L. Amico, L. Fazio, A. Osterloh, V. Vedral,
{\it Rev. Mod. Phys.} {\bf 80}, 517 (2008).

\bibitem{TMB10} M. Tichy, F. Mintert, A. Buchleitner,
{\it J. Phys. B: At. Mol. Opt. Phys.} {\bf 44}, 192001 (2011).




\bibitem{AM04} C. Amovilli, N. H. March,
{\it Phys. Rev. A} {\bf 69}, 054302 (2004).

\bibitem{OS07} O. Osenda, P. Serra,
{\it Phys. Rev. A} {\bf 75}, 042331 (2007).

\bibitem{OS08} O. Osenda, P. Serra
{\it J. Phys. B: At. Mol. Opt. Phys.}
\textbf{41}, 065502  (2008).

\bibitem{DKYPE2012} J. S. Dehesa, T. Koga, R. J. Ya\~nez, A. R. Plastino, R. O. Esquivel,
{\it J. Phys. B: At. Mol. Opt. Phys.} {\bf 45}, 015504 (2012).


\bibitem{CSD08} J. P. Coe, A. Sudbery, I. D'Amico,
 {\it Phys. Rev. B} {\bf 77}, 205122 (2008).

\bibitem{PN09} J. Pipek, I. Nagy,
{\it Phys. Rev. A} {\bf 79}, 052501 (2009).


\bibitem{HF11} N. L. Harshman, W. F. Flynn, {\it Quant. Inf. Comp.} {\bf 11}, 278 (2011).

\bibitem{KO10} P. Ko\'scik, A. Okopi\'nska,
{\it Phys. Lett. A} {\bf 374}, 3841 (2010).

\bibitem{K11} P. Ko\'scik, {\it Phys. Lett. A} {\bf 375}, 458 (2011).

\bibitem{KH2012} P. Ko\'scik, H. Hassanabadi,
\newblock {\it Few-Body Systems} {\bf 52}, 189 (2012).

\bibitem{KO13}
P. Ko\'scik, A. Okopi\'nska,
\newblock {\it Few-Body Systems} {\bf 54}, 1637 (2013).

\bibitem{KO14}
P. Ko\'scik, A. Okopi\'nska,
\newblock {\it Few-Body Systems} {\bf 55}, 1151 (2014).

\bibitem{GN05} M. L. Glasser, L. M. Nieto,
{\it J. Phys. A: Math. Theor.} {\bf 38}, L455 (2005).

\bibitem{MPD2012}
A. P. Majtey, A. R. Plastino, J. S. Dehesa,
\newblock {\it J. Phys. A: Math. Theor.} {\bf 45}, 115309 (2012).

\bibitem{NSPC12} R. G. Nazmitdinov, N. S. Simonovic, A. R. Plastino,
A. V. Chizhov,
{\it J. Phys. B: Atom. Mol. Opt. Phys.} {\bf 45}, 205503 (2012).

\bibitem{SB12}
P. Sadhukhan, S. M. Bhattacharjee,
{\it J. Phys. A: Math. Theor.} {\bf 45}, 425302 (2012).

\bibitem{LH13}
Y. C. Lin, C. Y. Lin, Y. K. Ho,
{\it Phys. Rev. A} {\bf 87},  022316 (2013).

\bibitem{SFM13}
S. Schroter, H. Friedrich, J. Madronero,
{\it Phys. Rev. A} {\bf 87}, 042507 (2013).

\bibitem{LLH13}
C. H. Lin,  Y. C.Lin, Y. K. Ho,
{\it Few-Body Systems} {\bf 54}, 2147 (2013).

\bibitem{RS14}
J. P. Restrepo, J. L. Sanz-Vicario,
{\it J.  Phys. C Conference Series} {\bf 488}, 152004 (2014).

\bibitem{RS2015} J. P. Restrepo Cuartas and J. L. Sanz-Vicario
Phys. Rev. A {\bf 91}, 052301 (2015).

\bibitem{LH14}
C. H. Lin, Y. K. Ho,
{\it Phys. Lett. A} {\bf 378}, 2861 (2014)




\bibitem{NV07} J. Naudts, T. Verhulst, {\it Phys. Rev. A} {\bf 75}, 062104 (2007).

\bibitem{PMD09} A. R. Plastino, D. Manzano, J. S. Dehesa,  {\it Europhys. Lett.} {\bf 86}, 20005 (2009).


\bibitem{ESBL02} K. Eckert, J. Schliemann, D. Bruss, M. Lewenstein, {\it Ann. Phys. (N.Y.)} {\bf 299}, 88 (2002).

\bibitem{GM04} G. C. Ghirardi,  and L. Marinatto, {\it Phys. Rev. A} {\bf 70}, 012109 (2004).

\bibitem{GMW02} G. C. Ghirardi, L. Marinatto, T. Weber, {\it J.Stat.
Phys.} {\bf 108}, 48 (2002).

\bibitem{OSTS08} V. C. G. Oliveira, H. A. B. Santos, L. A. M. Torres, A. M. C. Souza, {\it Int. J. Quant. Inf.} {\bf 6}, 379 (2008).

\bibitem{BPCP08} A. Borras, A. R. Plastino, M. Casas, A. R. Plastino, {\it Phys. Rev. A} {\bf 78}, 052104 (2008).

\bibitem{ZP10} C. Zander, A. R. Plastino, {\it Phys. Rev. A} {\bf 81}, 062128 (2010).




\bibitem{BR09} M. B. Ruiz,
{\it J. Math. Chem.} {\bf 46}, 24 (2009).

\bibitem{calais} J. L. Calais and P. O. L\"owdin, {\it J. Mol. Spectr.} {\bf 8}, 203 (1962).

\bibitem{cohl} H. S. Cohl,
{\it Integ. Transf. Spec. Funct.} {\bf 24}, 807 (2013).


\bibitem{rompabjes} E. Romera, P. S\'anchez-Moreno,  J. S. Dehesa, {\it J. Math. Phys.} {\bf 47}, 103504 (2006).

\bibitem{dehros}  J. S. Dehesa, S. L\'opez-Rosa, B. Olmos, R. J. Y\'a\~nez, {\it J. Math. Phys.} {\bf 47}, 052104 (2006).

\bibitem{avery} J. Avery, {\it Hyperspherical Harmonics and Generalized Sturmians} (Kluwer: Dordrecht, 2000).

\bibitem{olver} F. W. J. Olver {\it et al.}, {\it NIST Handbook of Mathematical Functions}, (Cambridge University Press
Cambridge, 2010).




\bibitem{BT1988} J.D. Barrow and F.J. Tipler, {\it The Anthropic Cosmological Principle} (Cambridge University Press, 1988).


\bibitem{Y82}
L.G. Yaffe, {\it Rev. Mod. Phys.} {\bf 54}, 407 (1982).


\end{thebibliography}
\end{document}